\documentclass[twocolumn,times]{aastex63}

\newcommand{\tess}{{\em TESS}}
\newcommand{\msun}{$M_{\odot}$}
\newcommand{\teff}{${T}_{\mathrm{eff}}$}
\newcommand{\logg}{$\log{g}$}
\newcommand{\kms}{km~s$^{-1}$}

\shorttitle{TESS-discovered companion to BPM 36430}
\shortauthors{Smith, Barlow, Rosenthal, Hermes \&  Schaffenroth}

\begin{document}


\title{Pulse Timing Discovery of a Three-Day Companion to the Hot Subdwarf BPM 36430}

\correspondingauthor{Bryce~A.~Smith}
\email{bsmith6@highpoint.edu}

\author[0000-0002-5717-2538]{Bryce~A.~Smith}\affil{Department of Physics and Astronomy, High Point University, High Point, NC, 27268, USA}
\author[0000-0002-8558-4353]{Brad~N.Barlow}\affil{Department of Physics and Astronomy, High Point University, High Point, NC, 27268, USA}
\author[0000-0001-9526-4436]{Benjamin Rosenthal}\affil{Friends Seminary, 222 E16th Street, New York, NY 10003, USA}\affil{Department of Astronomy, Boston University, 725 Commonwealth Ave., Boston, MA 02215, USA}
\author[0000-0001-5941-2286]{J.~J.~Hermes}\affil{Department of Astronomy, Boston University, 725 Commonwealth Ave., Boston, MA 02215, USA} 
\author[0000-0001-6339-6768]{Veronika ~Schaffenroth}\affil{Institute for Physics and Astronomy, University of Potsdam, Karl-Liebknecht-Str. 24/25, 14476 Potsdam, Germany}

\begin{abstract}
Hot subdwarf B stars are core-helium burning objects that have undergone envelope stripping, likely by a binary companion.  Using high-speed photometry from the {\em Transiting Exoplanet Survey Satellite}, we have discovered the hot subdwarf BPM\,36430 is a hybrid sdBV$_{\mathrm{rs}}$ pulsator exhibiting several low-amplitude $g$-modes and a strong $p$-mode pulsation. The latter shows a clear, periodic variation in its pulse arrival times.  Fits to this phase oscillation imply BPM\,36430 orbits a barycenter approximately 10 light-seconds away once every 3.1 d. Using the CHIRON echelle spectrograph on the CTIO 1.5-m telescope, we confirm the reflex motion by detecting a radial velocity variation with semi-amplitude, period, and phase in agreement with the pulse timings. We conclude that a white dwarf companion with minimum mass of $\approx$\,0.42\,$M_{\odot}$ orbits BPM\,36430. Our study represents only the second time a companion orbiting a pulsating hot subdwarf or white dwarf has been detected from pulse timings and confirmed with radial velocities.  
\end{abstract}


\section{Introduction} 
\label{sec:intro}

Hot subdwarf B (sdB) stars are evolved, low-mass objects believed to have helium-burning cores and thin hydrogen atmospheres. Their properties place them on the far blue end of the horizontal branch, known as the extreme horizontal branch. Most sdB stars have masses around 0.47\,\msun\ \citep{fontaine12}. Their effective temperatures range from \teff\ = $20{,}000-40{,}000$\,K, and their surface gravities range from \logg\ = $4.5-6.5$ \citep{heb16}. 

The red giant progenitors of hot subdwarfs experienced significant mass loss near the tip of the red giant branch. The majority of their hydrogen envelope was expelled, leaving behind only the helium-burning core and a thin hydrogen envelope. This hydrogen layer is too thin to sustain nuclear burning, and so hot subdwarfs will directly enter the white dwarf cooling sequence upon core helium exhaustion. \citet{han02,han03} proposed several Roche-lobe-overflow (RLOF) and common-envelope (CE) evolution channels that can produce hot subdwarfs through binary interactions. \citet{pel20} recently presented strong observational evidence that {\em all} hot subdwarf B stars must have been formed in binary systems. In most cases, the companions survive the RLOF and CE interactions, and studying their properties can give us important insight into the formation channels of these unique systems and help tune model parameters like envelope-binding energy, common-envelope-ejection efficiency, and angular momentum transfer (e.g., \citealt{sch22}). Hot subdwarf binaries are most commonly found from radial-velocity variations or flux changes in their light curves caused by eclipses, the reflection effect, ellipsoidal modulations, and/or Doppler beaming (e.g., \citealt{barlow21}).

Some hot subdwarf B stars exhibit pulsations, and they can be classified into three basic groups: slow gravity-mode ($g$-mode) pulsators (sdBV$_{\mathrm{s}}$ stars), rapid pressure-mode ($p$-mode) pulsators (sdBV$_{\mathrm{r}}$ stars), and hybrid pulsators (sdBV$_{\mathrm{rs}}$ stars).  \citet{lyn21} presented an overview of their properties and efforts to analyze them with asteroseismology.

Some sdBV pulsations are strong and stable enough to be used as precise ticks of a clock. Measuring their arrival times and comparing them to an ephemeris of predicted times allows one to constrain secular evolution rates of the star or look for signs of orbital reflex motion due to a nearby companion \citep{hermes18}. In the case of orbital reflex motion, pulses will be delayed when the sdBV is on the far side of its orbit, and they will be advanced when the sdBV is on the near side. The orbital period, radial velocity semi-amplitude, and minimum mass of the companion can be determined through precise measurements of these timing variations. \cite{bar11} and \cite{ota18}  used this method to find previously unknown companions to sdBV$_{\rm r}$ stars. The former was the first and only time that pulse timing results were confirmed using radial-velocity measurements for a compact star. 

BPM\,36430 (Gaia\,EDR3\,5371215147518355328; $G$ = 12.8\,mag; TIC 273218137) is a newly discovered sdBV$_{\rm rs}$ star displaying weak $g$-mode pulsations and a strong, radial-mode pulsation that is well-suited for pulse timing studies \citep{krz22}. Using \tess\ photometry, we measured precise pulse arrival times of the dominant 342-s pulsation period. We also monitored the radial velocities of BPM\,36430 with the CHIRON echelle spectrograph. Here, we present both sets of observations and show that BPM\,36430 displays orbital reflex motion every 3.1\,d due to a nearby companion.



\section{Time-Series Photometry} \label{sec:photometry}

\subsection{\tess\ Observations}
The {\em Transiting Exoplanet Survey Satellite} (\tess) provides extended time-series photometry for millions of objects across the entire sky \citep{ric14}. BPM\,36430 was observed by \tess\ in Sector 10 (at 2-min cadence) and Sector 37 (at 20-s cadence) through Guest Investigator Programs G011113 and G03221, respectively. The Sector 37 data are also available in their stacked 2-min cadence form.  
We used the Mikulski Archive for Space Telescopes\footnote{\href{https://archive.stsci.edu/missions-and-data/tess}{https://archive.stsci.edu/missions-and-data/tess}} (MAST) to download
the calibrated light curves, which were automatically reduced and corrected for instrumental systematics using the 
\tess\ data processing pipeline\footnote{\href{https://heasarc.gsfc.nasa.gov/docs/tess/pipeline.html}{https://heasarc.gsfc.nasa.gov/docs/tess/pipeline.html}} \citep{jen16}. For the flux we used the \texttt{PDCSAP\_FLUX} values, which are simple aperture photometry (\texttt{SAP\_FLUX}) values
corrected for systematic trends common to all stars on that chip. The Sector 10 and 37 light curves of BPM\,36430 have \texttt{CROWDSAP} values of 0.33 and 0.31, respectively and are moderately contaminated with light from other stars in the extraction aperture. Consequently, the measured amplitudes of any pulsations are diluted by background light. This dilution should not affect the results of this study, which relies on frequency and phase measurements.

\begin{figure}[t]
\centering
\includegraphics[width=0.975\columnwidth]{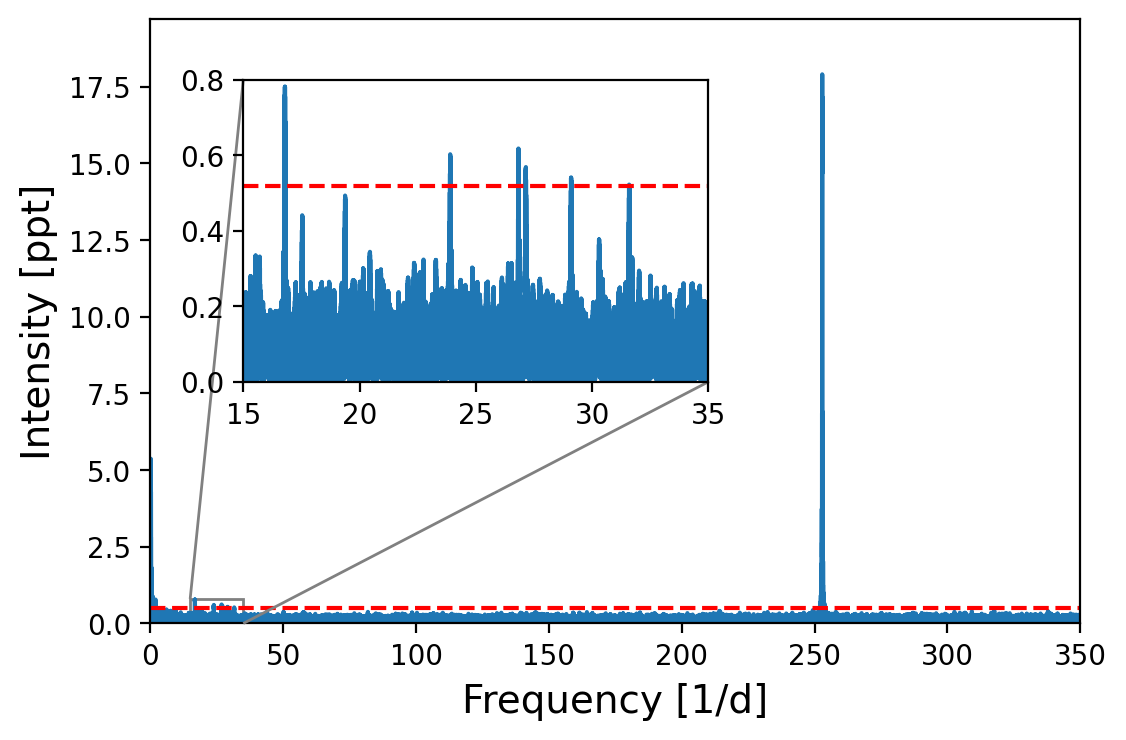}
\includegraphics[width=1.0\columnwidth]{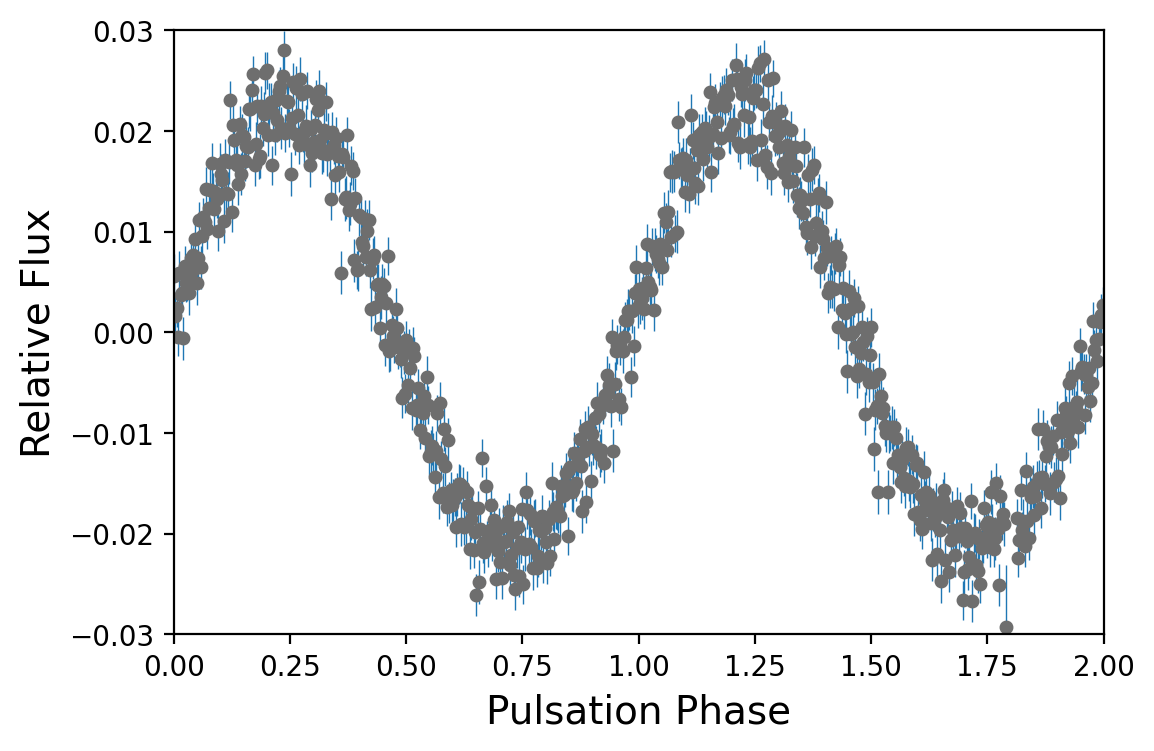}
\caption{Lomb-Scargle periodogram of the full, 2-min-cadence \tess\ light curve of BPM\,36430 showing a strong radial-mode pulsation and an inset of $g$-mode pulsations (top). A dashed red line denotes the 5.1-$\sigma$ level, which corresponds to a false alarm probability of 0.1\%. Also shown is the 20-s cadence light curve phase-folded on the dominant pulsation mode at 341.6734767\,s (bottom).}
\label{fig:LC_w_LSP_full}
\end{figure}

We calculated the Lomb-Scargle periodogram of both the 20-s and 2-min cadence light curves using the SciPy library \citep{vir20}. As shown in the top panel of Figure \ref{fig:LC_w_LSP_full}, BPM\,36430 is a hybrid sdBV$_{\rm rs}$ star showing both $g$- and $p$-mode pulsations. We used an iterative pre-whitening process to locate and measure all pulsation peaks. Table~\ref{tab:pulsations} presents the results of this analysis using the combined 2-min cadence light curve. We find four peaks consistent with $g$-mode pulsations that have amplitudes above the 5.1-$\sigma$ threshold, which corresponds to a false alarm probability (FAP) of 0.1\% \citep{bar21}. Additionally, we find three other $g$-modes that are below this threshold but appear above 4-$\sigma$ in each of the Sector\,10 and Sector\,37 light curve periodograms, when analyzed individually. Altogether, we report seven independent frequencies consistent with $g$-mode pulsations, all of which have amplitudes between roughly $0.45-0.75$\,ppt. 

The overall variability of BPM\,36430 is dominated by a single, 342-s oscillation mode with amplitude around 2.2\%\footnote{Due to phase smearing, the amplitude of the 342-s pulsation appears lower in the 2-min cadence \tess\ light curve.}.  Although we cannot provide definitive classification of the pulsation mode, we note that its period and large amplitude are quite similar to known radial p-modes in other sdBV pulsators like CS\,1246 \citep{bar10} and Balloon\,090100001 \citep{bara08}.


\begin{table}[ht]
\centering
\caption{Pulsation Frequencies from 2-min Cadence Data}
\begin{tabular}[t]{crrl}
\hline
\hline
Frequency & \multicolumn{2}{c}{ Amplitude} & Comments\\ 

[$d^{-1}$] & [ppt] & [$\sigma$]& \\ 
\hline
16.791015(60) & 0.75(11) & 6.9 & $g$-mode\\ 
17.535(28) & 0.46(12) & 4.2 & $g$-mode\\ 
19.378(33) & 0.53(13) & 4.9 & $g$-mode \\ 
23.8973(21) & 0.66(12) & 6.0 & $g$-mode\\ 
26.8361(37) & 0.60(14) & 5.6 & $g$-mode \\ 
29.098(28) & 0.65(11) & 6.0 & $g$-mode \\ 
31.596(76) & 0.53(12) & 4.9 & $g$-mode\\ 
252.8730086(26) & 18.61(13) & 171.0 & $p$-mode, likely radial\\
\hline
\end{tabular}
\label{tab:pulsations}
\end{table}

\vspace{0.5cm} 

\subsection{O--C Diagram}

\begin{figure}[t]
\centering
\includegraphics[width=1.0\columnwidth]{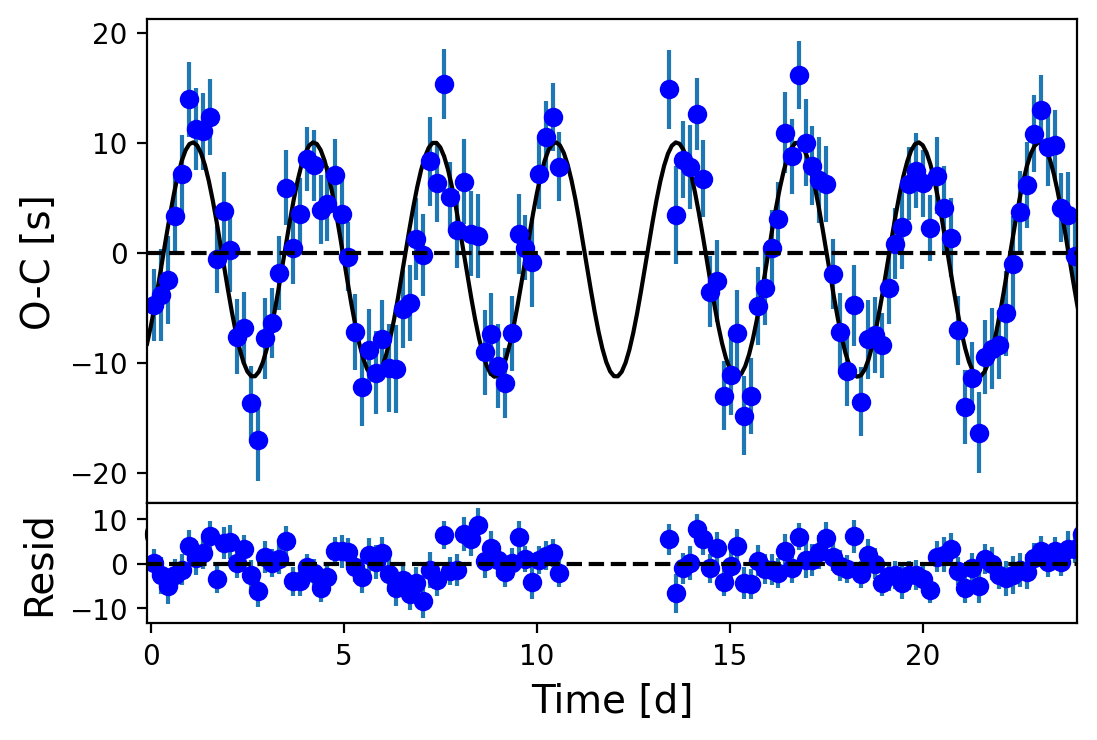}
\caption{$O-C$ diagram (top) constructed from pulse timing measurements of roughly 0.2-d subsets of the 342-s pulsation period, along with residuals after subtracting the best-fitting sinusoid from the data (bottom).}
\label{fig:O-C}
\end{figure}

BPM\,36430 is well-suited for pulse timing analysis due to its simple frequency spectrum being dominated by a single pulsation.
In order to quantify the stability of the main pulsation mode, we divided the 20-s \tess\ light curve into subsets using a K-means clustering algorithm\footnote{\href{https://github.com/dstein64/kmeans1d}{https://github.com/dstein64/kmeans1d}} and targeted each epoch to be roughly 0.2-d long.  Data subsets were generated so that none spanned the large download gap in the middle of the sector. We did not include the 2-min cadence light curve in this analysis given its slower sampling rate, which would lead to less reliable timing due to phase smearing. We used linear least-squares regressions to measure the pulsation phase within each subset of the 20-s cadence light curve. We plot relative phase shifts in an observed-minus-calculated ($O-C$) diagram, shown in Figure~\ref{fig:O-C}. 

The pulse timings of BPM\,36430 exhibit a strong, sinusoidal phase oscillation once every 3.1\,d, with a semi-amplitude of around 10\,s. Although some sdBV stars do show intrinsic amplitude and even phase modulation in their pulsations \citep{kil10,zong18}, the simplest explanation for the observed phase oscillation is reflex motion of the sdB due to a nearby companion in a circular orbit. In this case, the $O-C$ diagram reveals light-travel-time variations in the pulsations as the sdBV orbits the barycenter of the binary. When the sdBV is on the side of its orbit closest to Earth, we detect its pulses 10\,s earlier than we do on average, and similarly we detect a 10-s delay when it is on the far side of its orbit.

\subsection{Pulse Timing Analysis}

\begin{figure}[t]
\centering
\includegraphics[width=1.0\columnwidth]{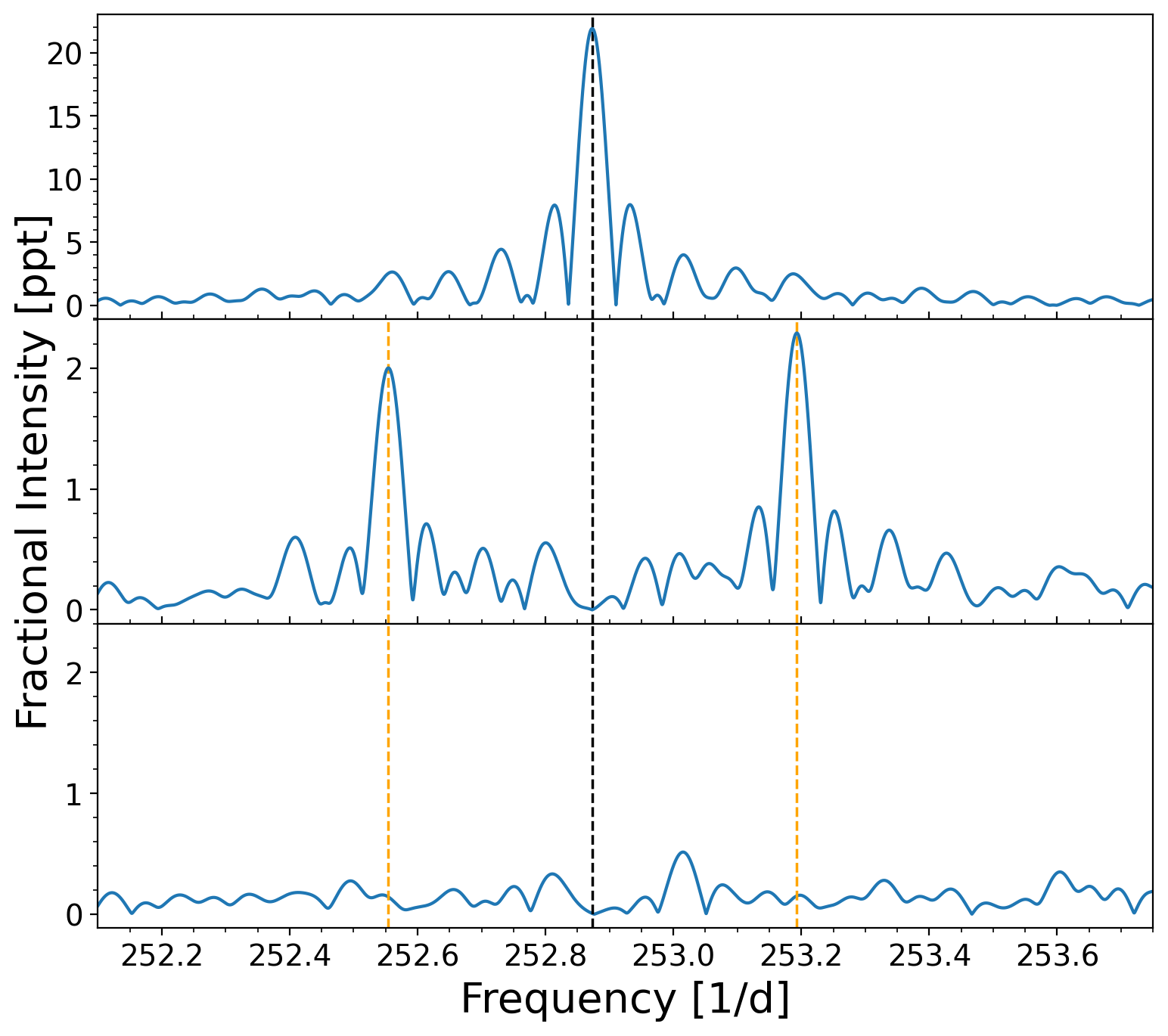}
\caption{Lomb-Scargle periodogram of the 20-s cadence \tess\ light curve of BPM\,36430, zoomed in on the dominant pulsation mode (top). Also shown are the periodograms after subtracting a fixed sinusoid (middle) and a phase-modulated sinusoid from the data (bottom). The dashed vertical lines mark position of the pulsation mode (black) and the two equidistant splittings (orange) arising from the orbital frequency of 0.32 cycles per day.}
\label{fig:PDGRM}
\end{figure}

\begin{table*}[t]
\centering
\caption{Best-Fitting Parameters to Equation \ref{eq:Pv_sin} from Photometry}
\begin{tabular}[t]{crlrlll}
\hline
\hline
Parameter & \multicolumn{2}{c}{Value from 20-s Data} & \multicolumn{2}{c}{Value from 2-min Data} & Unit & Comments\\
\hline
$A_p$& 2.213 & $\pm$ 0.013 & 1.88 & $\pm$ 0.10 & \%& pulsation semi-amplitude$^a$ \\
$P_p$& 341.673& $\pm$ 0.002 & 341.673475 & $\pm$ 0.000003 & s& pulsation period\\
$\phi_{p}$& 5.126 & $\pm$ 0.006 & 5.114 & $\pm$ 0.008 & radians & pulsation phase$^b$\\
$A_{b}$& 10.7 & $\pm$ 0.5 & 9.7 & $\pm$ 0.4 &  s& phase oscillation semi-amplitude\\
$P_{b}$& 3.132 & $\pm$ 0.009 & 3.12465 & $\pm$ 0.00018 & d & phase oscillation period\\
$\phi_{b}$& 1.52& $\pm$ 0.04 & 1.58 & $\pm$ 0.06 & radians & phase oscillation phase$^b$\\
\hline
\multicolumn{7}{l}{$^a$pulsation amplitude in 2-min-cadence data suffers phase smearing}\\
\multicolumn{7}{l}{$^b$phase at BJD$_{\rm TDB}$ = 2459320.485462}
\end{tabular}
\label{tab:phot_param}
\end{table*}

Since the \tess\ light curve is essentially continuous with a simple window function, we can model the entire light curve simultaneously with a phase-dependent sinusoidal function, instead of dividing the data into chunks to analyze the pulsation phases and generate an $O-C$ diagram. As previously discussed, the Lomb-Scargle periodogram of the \tess\ light curve shows a strong pulsation at 342\,s (top panel Figure~\ref{fig:PDGRM}). The phase oscillations found in the $O-C$ analysis should be encoded in the periodogram as a splitting of the main pulsation peak.

To look for and quantify this feature, we subtracted from the full light curve the best-fitting sinusoid with constant amplitude, period, and phase. Subtracting the main pulsation peak revealed two smaller peaks equidistant from the original pulsation signal (middle panel Figure~\ref{fig:PDGRM}). While such peaks could be indicative of rotational splitting, the $O-C$ diagram already gives us reason to believe there is a phase oscillation inherent in the pulsation. Therefore, we assume that the equidistant splittings encode this phase oscillation. We fitted a phase-modulated sinusoid to the data to determine the pulsational and orbital parameters of the potential binary system simultaneously using the expression:


\begin{equation}
f(t) = A_p\, sin \left(\frac{2 \pi t}{P_p} + \phi(t) \right)
\label{eq:Pv_sin}
\end{equation}

\noindent where  $\phi(t)$, the new time-variable phase due to the orbital reflex motion, is given by 

\begin{equation}
\phi(t) = \phi_p +2\pi\left(\frac{A_b}{P_p}\right)sin\left(\frac{2 \pi t}{P_b}+\phi_b\right) 
\end{equation}

\noindent where $A_b$ is the light travel time across the radius of the orbit of BPM\,36430, $P_b$ is the period of the orbit, and $\phi_p$ and $\phi_b$ are phases for the pulsation and binary orbit, respectively. When this phase-modulated form of the sinusoid was subtracted from the full light curve, only noise remained (bottom panel Figure~\ref{fig:PDGRM}).  We carried out the above analysis on both the 20-s cadence light curve (Sector 37) and the 2-min cadence light curve (Sectors 10 $+$ 37).

We ran Markov Chain Monte Carlo (MCMC) iterations to determine parameter uncertainties using \texttt{emcee}, an implementation of an MCMC sampler that uses a number of parallel chains to explore the solution space. Our MCMC sampling used 100,000 steps with 99 walkers \citep{for19}. Once the distribution was sampled, the best-fitting parameters and their uncertainties were determined. Our MCMC corner plots all have normally distributed posterior distributions, indicating that we did not have significant covariances in our fit.

As shown in Table~\ref{tab:phot_param}, the results from fitting the 20-s and 2-min cadence light curves agree with one another within the uncertainties. The only exception is the pulsation amplitude, which, due to phase smearing, is reduced by 19.1\% and 0.6\% in the 2-min and 20-sec light curves respectively. After inflating these amplitudes by factors of 1.24 and 1.006 to correct for smearing \citep{baldry99}, we derive consistent amplitudes of $A_p=2.33 \pm 0.12 \%$ and $A_p=2.226 \pm 0.013 \%$ for the 2-min and 2-sec light curves respectively. 

If the variation in the arrival times of the pulses is interpreted as circular reflex motion, the phase oscillation period and strength give us the orbital parameters. Adopting the more precise values from the 2-min light curve analysis, we measure the period and radius of the sdB orbit to be $3.12465 \pm 0.00018$\,d and $9.7 \pm 0.4$\,light-seconds, respectively. It is possible that the star could exhibit internal variations leading to oscillatory pulse-timing variations without the need for an orbiting companion (e.g., \citealt{dalessio13,zong16}). To confirm binarity, radial-velocity monitoring is required. Under the assumption of orbital reflex motion, our pulse-timing results predict that the hot subdwarf should show a $K_{\rm sdB} = 67.7 \pm 2.9$\,\kms\ radial-velocity variation with a 3.1-d period. 

\section{Time-Resolved Spectroscopy}
\label{sec:spec}

\subsection{CHIRON Observations}

In order to confirm orbital reflex motion as the cause of the pulse-arrival-time variations, we obtained 25 high-resolution spectra ($R=28{,}000$) of BPM\,36430 in 2021 and 2022 using the fiber-fed CHIRON echelle spectrograph on the 1.5-m SMARTS telescope at the Cerro Tololo Inter-American Observatory \citep{tok13}. All spectra covered the wavelength range $4500-8360$\,\AA\ and were exposed for exactly 1367\,s, or four cycles of the 342-s pulsation. This specific integration time was chosen to insure that the physical expansion and contraction of the star during each pulse did not influence our radial-velocity measurements. Data were reduced using a customized reduction pipeline described in \citet{tok13}, which is run by members of the RECONS team\footnote{\href{http://recons.org/}{http://recons.org/}} at Georgia State University.

\subsection{Radial Velocity Curve}
\begin{figure}[t]
\centering
\includegraphics[width=1.0\columnwidth]{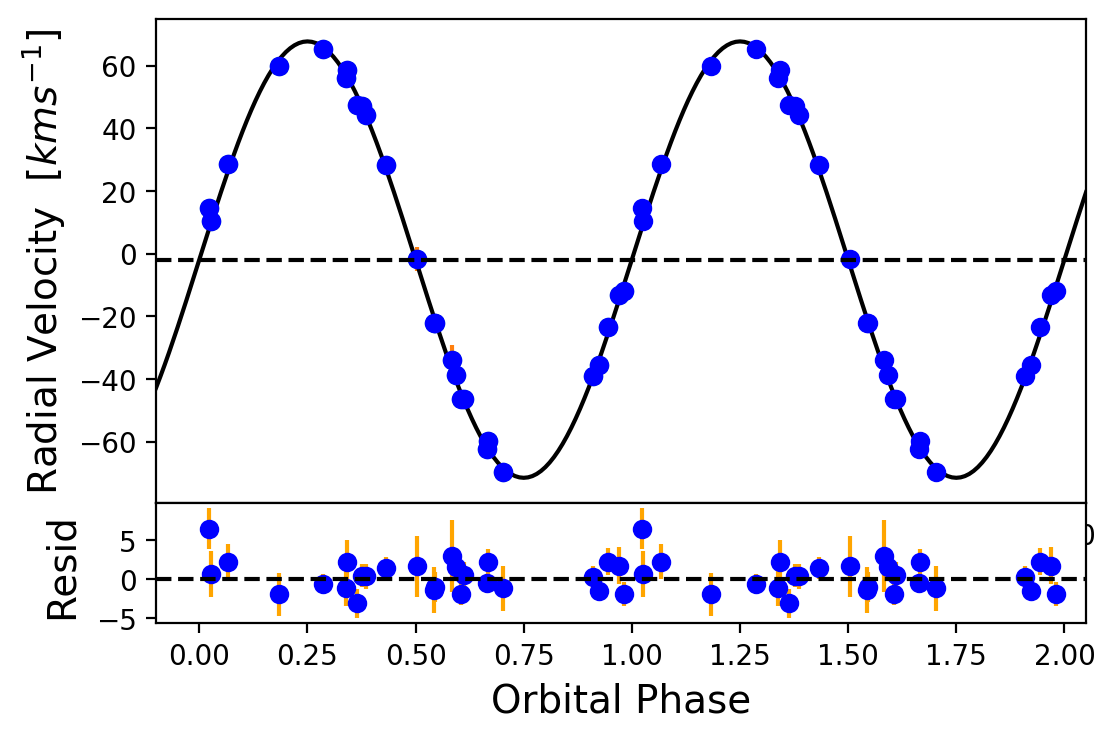}
\caption{Radial-velocity curve of the sdBV primary in BPM\,36430 (top). Residuals after subtracting the best-fitting sinusoid are also shown (bottom).}
\label{fig:RV_PF}
\end{figure}

The He-I absorption line at 5875\,\AA\ was the only spectral feature with high enough signal-to-noise to use for measuring radial velocities. While H$\alpha$ and H$\beta$ were also visible, their profiles each spanned multiple echelle orders, and so they did not produce reliable results. We wrote a custom Python code that used SciPy's \texttt{curve\_fit} routine \citep{vir20} to fit inverse Gaussians to the absorption profiles. Individual radial velocities were calculated using the Doppler shift from the accepted NIST\footnote{\href{http://physics.nist.gov}{http://physics.nist.gov}} value. The RV semi-amplitude of the sdB was determined by using the same \texttt{curve\_fit} function to fit a sine wave to the velocities. The RV curve demonstrated clear orbital reflex motion with period $P_{b} = 3.12458 \pm 0.00012$\,d and semi-amplitude of $K_{\rm sdB} = 69.6 \pm 0.6$\,\kms. These values agree with the results of our light curve analysis within their 1-$\sigma$ uncertainties. Figure~\ref{fig:RV_PF} presents the RV curve phase-folded over the orbital period, and Table~\ref{tab:rv} shows the best-fitting parameters. 


\begin{table}[t]
\centering
\caption{Best-Fitting Binary Parameters from Spectroscopy}
\begin{tabular}[t]{crlll}
\hline
\hline
Param & Value& Uncertainty & Unit& Comments\\
\hline
$P_b$& 3.12458& $\pm$ 0.00012& d& orbital period\\
$K_{\rm sdB}$& 69.6& $\pm$ 0.6& \kms\ & sdB RV semi-amp.\\
$\gamma$ & $-$1.9 & $\pm$ 0.4 & \kms\ & systemic velocity\\
$f$& 0.109& $\pm$ 0.003& \msun\ & binary mass func.\\
\hline
\end{tabular}
\label{tab:rv}
\end{table}

\section{Nature of the Hidden Companion} 
\label{sec:discussion}


In order to determine the nature of the companion in BPM\,36430, we must first confirm the status of the primary star as a hot subdwarf and constrain its mass. Unfortunately, we do not possess an identification spectrum to model with synthetic spectra, which is the easiest way to derive the effective temperature and surface gravity. However, we can still measure these parameters, along with the radius $R$ and luminosity $L$, by fitting the spectral energy distribution (SED) of the target with model spectra and combining the results of this fit with the distance constrained by its {\em Gaia} parallax. 
Full details of our SED fitting method are presented by \citet{heb18} and \citet{irr21}. An analysis of several well-studied sdB binaries with white dwarf and low-mass main sequence companions by \citet{sch22} shows that the SED determines effective temperature quite reliably for hot subdwarfs with \teff\ $<$ $32{,}000$\,K. 

\begin{figure}[t]
\centering
\includegraphics[width=1.0\columnwidth]{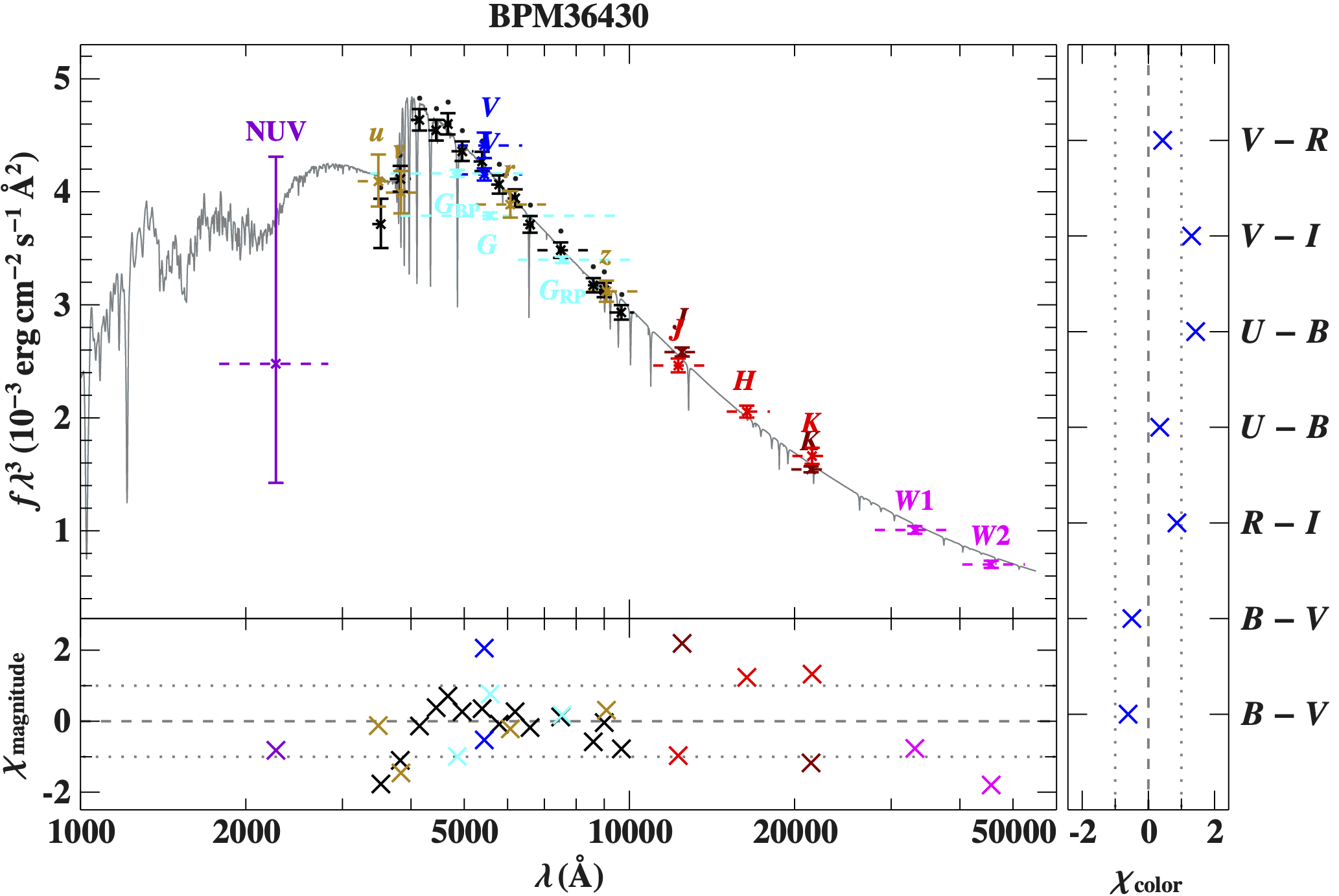}
\caption{Spectral energy distribution of BPM\,36430 and our model fit, which shows that a $\approx$29,000 K sdB star dominates the light from the system.}
\label{fig:SED}
\end{figure}

Figure~\ref{fig:SED} and Table~\ref{tab:sed} present the results of our SED fitting, which confirm that the primary star in BPM\,36430 is likely an sdB. We find a surface temperature \teff\ $\approx 29{,}400$\,K, radius of $R$ $\approx$ 0.2 $R_{\odot}$, and luminosity of $L$ $\approx$ 26 $L_{\odot}$. We are unable to calculate masses from these values since we lack a measured \logg\ from spectroscopy. However, we note that there are several sdBV stars that are dominated by a single pulsation mode quite similar in period and strength to that of BPM\,36430, and they all cluster tightly in the \logg\ - \teff\ plane, near \teff\ $\approx 29{,}000$\,K and \logg\ $\approx 5.45$. Such stars include CS 1246 \citep{bar10}, Balloon 090100001 \citep{ore04}, RAT0455+1305 \citep{baran10}, J08069+1527 \citep{baran11}, and five other unpublished sdBVs (Bradshaw et al., 2022, priv. comm.)  If we assume a typical surface gravity for these similar pulsating hot subdwarfs (log $g$ = 5.45), we derive an sdB mass near 0.40 \msun. This mass, combined with the orbital period $P_{b}$ and sdB RV semi-amplitude $K_{\rm sdB}$, results in a companion mass of $m \sin i$ $\approx$ 0.42\,\msun\ orbiting at a separation distance of $a \sin i \approx 8.4$\,$R_\odot$. Without eclipses, we do not know the orbital inclination angle and are limited to only computing minimum values for the separation distance and companion mass.

\begin{table}[t]
\label{tab:sed}
\centering
\caption{Best-Fitting Parameters from the SED Modeling}
\begin{tabular}[t]{lr}
\hline
\hline
Parameter & 68\% Conf. Interval\\
\hline
Color excess E(44-55) & 0.057$^{+0.008}_{-0.008}$ mag\\
Angular diameter log($\Theta$ (rad)) & $-$10.775$^{+0.010}_{-0.011}$\\
Parallax $\bar{\omega}$ (Gaia, RUWE = 0.80) & 1.88 $\pm$ 0.04 mas\\
\hline
Effective temperature T$_{\rm eff}$ & 29400$^{+1100}_{-900}$ K\\
Surface gravity log(g (cm s$^{-2}$)) (fixed) & 5.45\\
Radius R = $\Theta$/(2$\bar{\omega}$) & 0.198$^{+0.007}_{-0.006}$ $R_{\odot}$\\
Mass M = gR$^2$/G & 0.40 $\pm$ 0.03 M$_{\odot}$\\
Luminosity L/L$_{\odot}$  & 26$^{+5}_{-4}$\\
Gravitational redshift $v_{{\rm grav}}$ = GM/(Rc) & 1.29$^{+0.05}_{-0.05}$ \kms \\
\hline
\end{tabular}
\label{tab:sed}
\end{table}


There are several reasons that we believe the companion is a white dwarf. First, the period and companion mass are much more consistent with those of known sdB+WD systems. \citet{sch22} shows that our sdB and companion masses fall directly in the middle of the observed distributions for sdB+WD binaries, and that our 3.1-d orbital period is much longer than those observed in sdB+dM/BD binaries. Second, if the companion were a main-sequence star with a mass of 0.42\,\msun\ or greater, we would expect to see an infrared excess. Our SED shows no such excess and is consistent with a single sdB star (Figure~\ref{fig:SED}). Finally, if the companion star were a late-type main sequence star, we would expect to see a reflection effect as a signal in the periodogram, with peaks at both the orbital frequency and possibly its first harmonic. The amplitude at the orbital frequency would be $\approx 1$\%, several times above our detection limit (see Figure~15 from \citealt{sch22}). We see neither of these signals in the 2-min cadence or 20-s cadence light curve periodograms. 

With such a short period and separation distance, BPM\,36430 likely formed through the stable RLOF+CE channel presented by \citet{han02,han03}, if it is truly a sdB+WD binary. In this scenario, two main-sequence stars existed in a binary. The more massive star evolved to the red-giant stage first. It filled its Roche lobe, transferred mass, and then became the white dwarf. Then, the less-massive main-sequence star (the progenitor to the sdB we see today) evolved to become a red giant. It began unstable RLOF mass transfer to the white dwarf star. A common envelope formed, which brought the stars closer together. Once the common envelope was ejected, this left behind an sdB star with a white dwarf companion in a tight orbit. This model predicts that the sdB and companion would have similar masses, which is the case for a 0.40\,\msun\ sdB orbited every 3.1-d by a $>$0.42\,\msun\ companion \citep{han02,han03}.

\section{Summary} 
\label{sec:summary}
Using 2-min and 20-s cadence data from \tess, we have discovered hybrid $g$- and $p$-mode pulsations in the sdB star BPM\,36430. Its photometric variations are dominated by a single pulsation mode with period of 342\,s and amplitude around 2\%. From a pulse timing analysis, we find a clear, 3.1-d periodic variation in the arrival times of this signal. We interpreted the cause of this variation to be orbital reflex motion. To confirm this we obtained spectroscopic observations with the CHIRON echelle spectrograph and performed a radial-velocity analysis. The orbital period and RV semi-amplitude derived from the photometric phase delay via an $O-C$ analysis and our spectroscopic RV analysis are consistent within 1$\sigma$, confirming that the variation in the pulsation arrival timing is due to orbital reflex motion. Given the period of the system and lack of infrared excess, the companion is most likely a white dwarf.

Our results provide additional evidence of the practicality of using pulse-timing measurements to uncover companions to hot subdwarfs. It marks only the second time for which a hot subdwarf binary found via pulse timings has been confirmed using RV measurements. 
Uncovering additional similar systems with the pulse timing and other methods will help further constrain the binary population synthesis models producing these enigmatic objects. The sensitivity of the pulse timing method to low-mass companions brings the exciting opportunity to find substellar and even planetary companions to hot subdwarfs, although follow-up RV measurements are a necessity for confirming any such discoveries. 

\section*{Acknowledgements}
B.B. and B.S. acknowledge financial support from NASA Grant 80NSSC21K0364. J.J.H. acknowledges financial support from \tess\ Guest Investigator Programs 80NSSC20K0592 and 80NSSC22K0737. J.J.H. would also like to thank the Maria Mitchell Observatory, as well as coordination from the 2021 RISE program at Boston University. V.S. was funded by the Deutsche Forschungsgemeinschaft under grant GE2506/9-1. B.B. and B.S also acknowledge Berggasthof Eckbauer for providing good space to have discussions about this manuscript. B.S. would like to personally acknowledge Bean Foster's Cafe in Golden, CO, for providing a good space to work on this manuscript. B.S. and B.B. would also like to acknowledge yogurt, the potential existence of an additional collaborator, the Core Four, and one spilled side of ranch. 

This research made use of \texttt{Lightkurve}, a Python package for Kepler and \tess\ data analysis \citep{lk18}. This paper includes data collected by the \tess\ mission, which are publicly available from the Mikulski Archive for Space Telescopes (MAST). Funding for the \tess\ mission is provided by NASA's Science Mission directorate.



\end{document}